# Ultrafast annealing process of MTJ using hybrid microwave annealing

Ming-Chun Hsu, Fan-Yun Chiu, Wei-Chi Aeneas Hsu, Chang-Shan Shen, Kun-Ping Huang and Tsun-Hsu Chang*, Senior Member, IEEE

*Abstract*—This paper discovers that the magnetic tunnel junction (MTJ) structure is successfully magnetized with hybrid microwave annealing, confirmed by the tunneling magnetoresistance (TMR) and Coercivity (Hc) results. Hybrid microwave annealing can transform CoFeB into a single crystal and form the Fe-O bond at the interface between CoFeB and MgO without adding an extra magnet. The
annealing time is significantly reduced from the original 120 minutes to just 1 minute, allowing for rapid low-temperature annealing of the MTJ structure. The TEM results are used to determine the change in the lattice structure of CoFeB from amorphous to a single crystal, and the EELS result indicates the diffusion distribution of atoms in the MTJ structure. This hybrid annealing process can save a significant amount of fabrication time and is an energy-efficient alternative to the current fabrication process of MRAM.

## I. INTRODUCTION

Magnetic Tunnel Junction (MTJ) is a key element of STT-MRAM [1]. Due to its nonvolatile, high-speed (~5ns), high-density (~F2), and high-retention (~10 years) characteristics, STT-MRAM can replace SRAM [2] and apply to in-memory computing for artificial intelligence [3].

The greatest challenge in designing MTJs for commercial applications is achieving high tunneling magnetoresistance (TMR) and coercivity (Hc) values, representing the success rate of reading data and stability. In MTJs based on CoFeB/MgO/CoFeB, the crystalline quality determines the TMR value. Commonly, the MgO deposited by RF sputtering exhibits firm texture (001) on the as-deposited amorphous CoFeB layer, which serves as a standard layer for annealing CoFeB. High-temperature annealing (>500 °C) can cause metal diffusion within the MTJ, leading to deterioration in the quality of CoFeB. Therefore, annealing temperatures are typically kept from 250 °C to 500 °C, with annealing times lasting from 1 to 2 hours. The magnet also helps align the perpendicular magnetic anisotropy at the interface of CoFeB and MgO during the thermal annealing [4].

Ming-Chun Hsu, Wei-Chi Aeneas Hsu and Chang-Shan Shen are with the College of Semiconductor Research, National Tsing-Hua University, Hsinchu 300, Taiwan (email: enoke19031904@gmail.com).

Tsun-Hsu Chang is with the Department of Physics, National Tsing-Hua University, Hsinchu 300, Taiwan (email: thschang@phys.nthu.edu.tw).

Fan-Yun Chiu and Kun-Ping Huang are with the Mechanical and Mechatronics Systems Lab, Industrial Technology Research Institute (email: fychiu@itri.org.tw).

In selecting annealing methods, apart from traditional furnace tube annealing, microwave and hybrid microwave annealing can also transform semiconductors into single crystals [5]. Moreover, the microwave is combined with alternating electric and magnetic fields, which control magnetism in ferromagnetic films [6-9] because electric fields redistribute the electron charge density, inducing the change in magnetism [10]. Another advantage of microwave annealing is the low-temperature process, which can reduce dopant diffusion and shorten annealing time. According to the above results, microwave annealing is a promising method for transforming the crystal structure and aligning the magnetism in the MTJ structure. However, one limitation is that the material must inherently possess the ability to absorb microwaves [11]. Hybrid microwave annealing is an improved method of microwave annealing that combines the advantages of traditional thermal annealing and microwave annealing. By introducing a susceptor (absorber) into the sample, materials that are not naturally good at absorbing microwaves can have their dielectric properties altered by the heating from the susceptor, making them more receptive to microwaves [12]. This study compared the effects of microwave annealing, hybrid microwave annealing, and traditional thermal annealing. TMR and Hc are used as indicators of the effectiveness of annealing. The experimental results demonstrate that hybrid microwave annealing can achieve TMR values similar to traditional thermal annealing while reducing the annealing time to one-tenth.

*A. Microwave annealing*

The alternating electric field in microwaves interacts with materials and can be divided into thermal and nonthermal effects. Thermal effect includes ohmic conductive loss and dielectric loss. The alternating electric field causes electrons and ions in the material to generate electrical currents, which collide with the material's lattice, resulting in energy transfer [13]. Nonthermal effects occur when microwaves directly interact with atomic bonding, causing energy transfer. This effect can increase the regrowth rate in the annealing process [14].

*B. Hybrid microwave annealing*

Hybrid microwave annealing (HMA) is an improved version of microwave annealing. Compared to MWA, HMA incorporates a susceptor for heat transfer. When the susceptor absorbs microwaves, it generates thermal energy by ohmic loss. As the susceptor conducts the thermal energy to the sample, the rising temperature of the sample affects its dielectric properties and attenuation distance [15, 16].

## II. EXPERIMENT AND SIMULATION

The Electronics and Optoelectronics Research Laboratories (EOL) at the Industrial Technology Research Institute (ITRI) provided the MTJ samples sputtered by a high-vacuum magnetron sputtering system at room temperature. Fig. 1 shows the MTJ's structural diagram.

The coercivity value of the MTJ was measured by the vibrating sample magnetometer (VSM), which has a maximum magnetic field of 1 Tesla at room temperature. The TMR value was measured using a Current-in-Plane Tunneling (CIPT) measurement system.

The conditions of furnace annealing were 360 °C for 2 hours in an in-plane magnetic field of 1T.

The microwave annealing cavity is a WR340 waveguide with a 2.45 GHz magnetron as the wave source, with a power range of 2.5 kW to 6 kW. The susceptor is the p-type silicon with 20 Ω·cm.

Field-emission transmission electron microscope (FE-TEM) was used. Elemental distribution within each film layer was analyzed using an Electron Energy Loss Spectrometer (EELS, Model 965 QuantumER™) attached to the Cs-TEM.

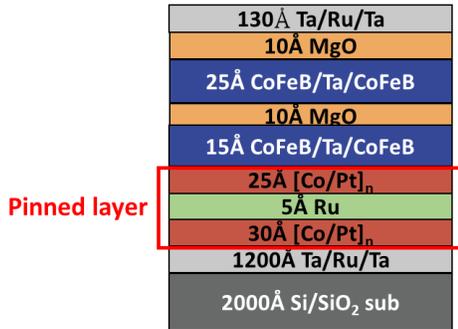

Fig. 1. Schematic Diagram of MTJ Material Structure

## III. RESULTS AND DISCUSSION

The TMR value of MTJ with different annealing processes is plotted in Fig.2. In this experiment, the MTJ samples underwent conventional furnace annealing at 360 °C for 120 minutes, resulting in a TMR value of 113%. The TMR value increased from 23 % to 30 % after microwave annealing, showing no significant annealing effect, indicating that the MTJ structure does not interact with microwaves. However, the TMR value in HMA increased from 39 % to 102 %. The susceptor's temperature altered the MTJ structure's dielectric properties, enabling microwave interaction. As a result, after just one minute of HMA, the TMR value approached the effect achieved by furnace annealing in two hours.

Fig. 3. shows the TMR and Hc values of the MTJ samples after HMA, using power levels ranging from 2.5 kW to 6 kW, each for one minute. Hc increased from 7 Oe at 2.5 kW to 35 Oe at 4kW, then decreased to 23 Oe at 4.5kW. At 4kW, the trend of Hc reversed, unlike TMR, which consistently followed the same trend. The reverse in Hc is usually caused by damage to the pinned layer, especially excessive intermixing in the Co/Pt multilayer [17], which disrupts interface anisotropy [18]. TMR is primarily generated by the interface anisotropy at the CoFeB/MgO interface, which is formed by Fe-O bonds [19]. Therefore, increasing the power in HMA can effectively facilitate the formation of Fe-O bonds. Based on the experimental results above, a power of 4 kW effectively enables the MTJ structure to achieve a TMR value greater than 80 % and an Hc value greater than 35 Oe after annealing. Therefore, subsequent experiments will fix the power at 4 kW while varying the annealing time.

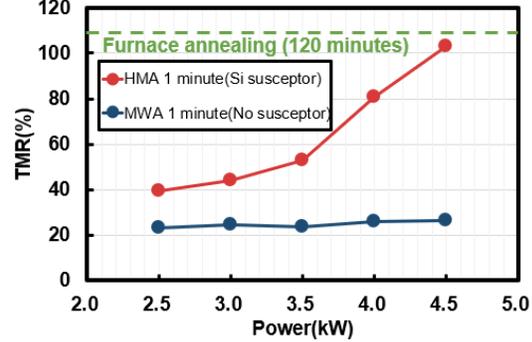

Fig. 2. TMR value of MTJ with different annealing processes

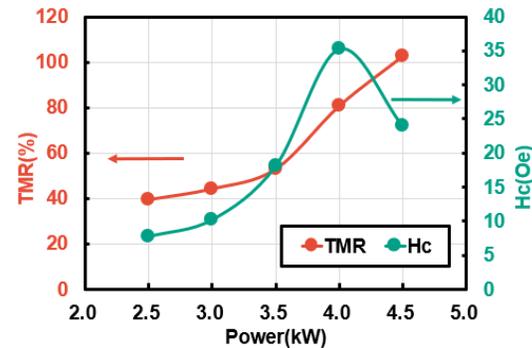

Fig. 3. TMR and Hc values of MTJ after HMA for 1 minute

Fig. 4. shows the TMR and Hc values of MTJ after HMA at 4kW, using the time range from half a minute to 2 minutes. The TMR value increased from 58 % to 96 %. The values for 1.5 minutes and 2 minutes could not be measured by the instrument, likely due to the destruction of the MTJ structure. Further confirmation will be conducted using TEM measurements. The Hc value slightly decreased from 43 Oe at 0.5 minutes to 40 Oe at 1 minute, while at 1.5 minutes, it dropped significantly to 4. This indicates that annealing time has a drastic impact on the pinned layer.

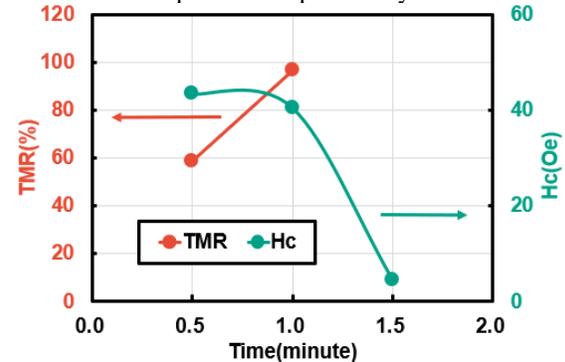

Fig. 4. TMR and Hc values of MTJ after HMA at 4kW

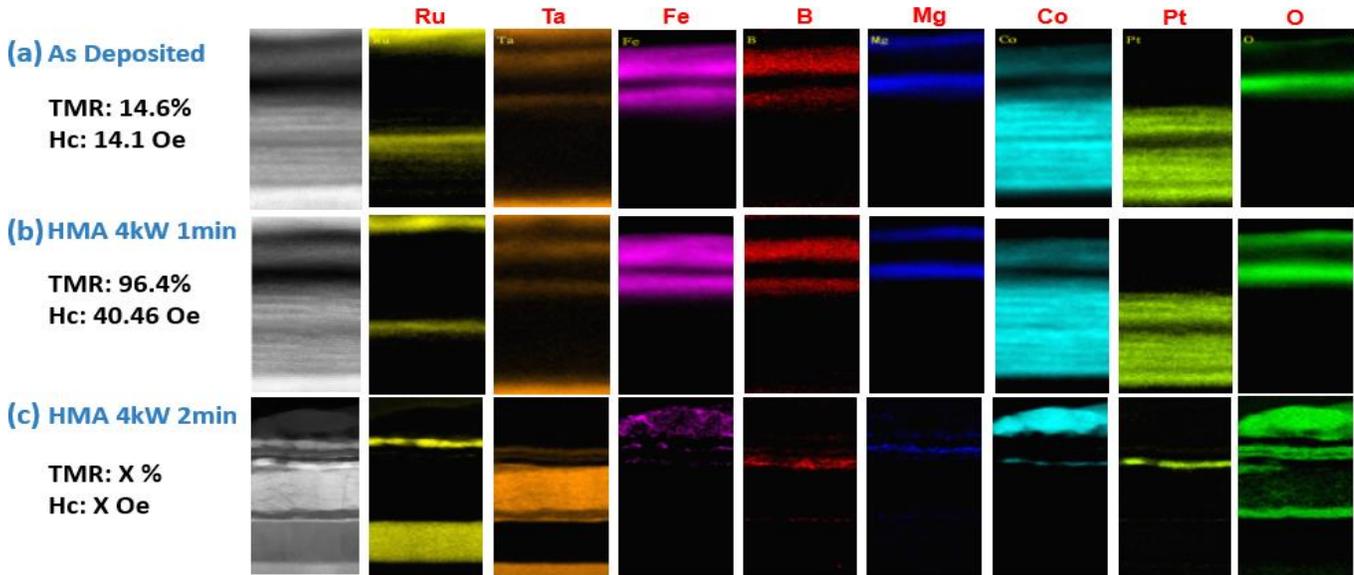

Fig. 5. A dark-field FE-TEM image and its corresponding electron energy-loss spectroscopy (EELS) mapping of the Ru, Ta, Fe, B, Mg, Co, Pt, O and  for (a) as-deposited and (b) annealed samples at 4kW for 1 minute (c) annealed samples at 4kW for 2 minutes

The dark-field FE-TEM image and its corresponding electron energy-loss spectroscopy (EELS) mapping are plotted in Fig. 5. At a power of 4 kW, the TMR increased from 14 % in the as-deposited state to 96 % after 1 minute of HMA. However, after 1.5 minutes of HMA, the TMR value could no longer be measured. TMR is formed by interface anisotropy generated from Fe-O bonding. As seen from the oxygen distribution map, after 1 minute of HMA, some oxygen diffused into the CoFeB layer, which bonded with Fe, explaining the increase in TMR to 96 %. However, after 2 minutes of HMA, the distribution of Fe and O had changed, preventing the formation of interface anisotropy and making TMR unmeasurable. Hc originates from the Co/Pt multilayer structure. The analysis shows that after 2 minutes of HMA, the Pt and Co multilayer structure was destroyed, making Hc undetectable.

## IV. Conclusion

The difference in TMR data between HMA and MWA demonstrates the improvement in the dielectric properties of the MTJ structure due to the susceptor. HMA enables the TMR value to approach that of furnace annealing while achieving a Hc of 40 Oe, ensuring sufficient thermal stability. Most importantly, the annealing time is significantly reduced from the original 120 minutes to just 1 minute, allowing for rapid low-temperature annealing of the MTJ structure. This means that both time and energy consumption can be reduced in wafer fabrication. Compared to furnace annealing, which heats the entire chamber, HMA enables targeted sample heating, significantly saving energy and reducing the carbon footprint.

The elemental distribution from EELS reveals the impact of HMA on the MTJ structure. Excessive annealing time causes interdiffusion within the multilayer structure of the pinned layer, leading to the loss of its original function. Additionally, Fe diffusion prevents oxygen from bonding with Fe to form Fe-O, resulting in insufficient interface anisotropy to generate TMR.

## Acknowledgment

Microsystems of National Tsing Hua University and Industrial Technology Research Institute provided technical support for VSM, CIPT, and TEM measurement.